\documentclass[12pt,preprint]{aastex}

\newcommand{\sixj}[6]{\ensuremath{\left\{{#1\atop #4}{#2\atop #5}
{#3\atop #6}\right\}}}

\begin{document}

\title{New Benchmark of X-ray Line Emission Models of \ion{Fe}{17}}
\author{
M. F. Gu}

\affil{Lawrence Livermore National Laboratory, 7000 East Avenue, Livermore, CA
  94550}

\begin{abstract}
We review the accuracy of existing \ion{Fe}{17} X-ray line emission models by
comparing them with an extensive analysis of \textit{Chandra} high energy
transmission grating (HETG) observations of stellar coronae. We find
significant discrepancies between most theoretical 
predictions and observations for at least some of the intensity ratios
involving the six principal Fe XVII lines, 3C (15.01~{\AA}),  
3D (15.26~{\AA}), 3E (15.45~{\AA}), 3F (16.78~{\AA}), 3G (17.05~{\AA}), and
M2 (17.10 {\AA}). We suggest that the main problem of most previous theoretical
studies to their inability to fully include electron correlation effects in
the atomic structure calculations, while any deficiencies in the scattering
approximation methods are of minor importance, regardless of it being
close-coupling (CC) or distorted-wave (DW). An approximate method based on the
many-body perturbation theory and DW approximation is proposed to include such
correlation effects in the calculation of collisional excitation cross
sections. The results are shown to agree with coronal observations and
laboratory measurements better than most previous theories. Using the new
atomic data, we then investigate the electron density sensitivity of the M2/3G
intensity ratio and provide an improved density diagnostic tool
for astrophysical observations. 
\end{abstract}


\section{Introduction}
There are several widely known problems in the theoretical modeling of
\ion{Fe}{17} X-ray emission lines, making it difficult to use them as reliable
diagnostic tools for electron density, temperature, plasma opacity, and iron
abundances in X-ray astronomy. The first problem is related to the intensity
ratio of the $2p_{3/2}$-$3d_{5/2}$ transition at 15.26~{\AA}, generally
referred to as 3D, to the $2p_{1/2}$-$3d_{3/2}$ transition, or 3C. Most
theoretical predictions of this ratio are smaller than astrophysical
observations \citep{xu02, brinkman00, canizares00, mewe01, behar01} and
laboratory measurements \citep{brown98, brown01a, beiersdorfer01}. This
discrepancy has been used in the past to infer plasma opacity as the stronger
3C line is more susceptible to resonant scattering \citep{waljeski94,
  saba99}. However, this diagnostics depends critically on knowing the
optically thin limit of the 3D/3C ratio. The second problem concerns the
intensity ratio of the $2p$-$3s$ (hereafter $3s$) transitions, including 3F
(16.78~{\AA}), 3G 
(17.05~{\AA}), and M2 (17.10~{\AA}), to the $2p$-$3d$ (hereafter $3d$)
transitions, including 
3C, 3D, and a much weaker 3E (15.45~{\AA}). The astrophysical observations and
laboratory measurements are also significantly larger than most theoretical
values \citep{phillips99, beiersdorfer02, beiersdorfer04} for this
ratio. Finally, the 
theoretical predictions of the intensity ratio of the two $2p$-$3s$
transitions, M2 to 3G, have also been shown to be significantly smaller than
astrophysical observations of low density plasmas \citep{ness05}. The M2/3G
ratio is density sensitive, and have been used to infer electron densities in
magnetic cataclysmic variables \citep{mauche01}. However, the inability of
reproducing the low density limit of this ratio in theory introduces
additional uncertainties in the density determination.

There have been claims that these modeling problems have been solved
with large scale close-coupling calculations of the electron collisional
excitation data of \citep{gchen02, gchen03}. The new theoretical values
of certain line ratios have been shown to agree with laboratory measurements
and a few astrophysical observations to within 10\%. However, the cross section
measurements of the 3C and 3D lines \citep{brown06} indicate that such
apparent agreements are rather fortuitous. The more recent work of
\citet{gchen07} and \citet{gchen08a} revised their earlier work by including
pseudo states and the $3d^2$ pair excitation configuration in the target
wavefunction expansion and obtained 15-20\% reductions in the 3C cross
sections compared with \citet{gchen03}, while those of 3D changed
little. \citet{gchen08b} further suggested that the theoretical radiative
recombination (RR) cross sections into the M-shell of \ion{Fe}{16} used for
normalization in \citet{brown06} are underestimated by 25\%, leading to
smaller measured 3C and 3D cross sections.
In this paper, we provide a
thorough assessment of the accuracies of existing atomic calculations on
\ion{Fe}{17} by comparing them with a large sample of \textit{Chandra}
observations of stellar coronae obtained with the high energy transmission
grating (HETG) spectrometer. We conclude that most of the existing predictions
do not offer satisfactory explanation of the observed line ratios. The recent
calculations of \citet{gchen07} and \citet{gchen08a} provide the best
agreement with the laboratory 3C and 3D cross sections. However, a detailed
comparison of these data with the \textit{Chandra} observations cannot be
done, as the full results of the calculations including all levels necessary
for collisional radiative modeling are not available. The major drawback of
\citet{gchen07} and \citet{gchen08a} is that it is not clear that the
accounting the target correlation effects has fully converged. In these
calculations, the largest improvement comes with the inclusion of $3d^2$ pair
excitation configuration. However, it is well know that such pair correlations are
notoriously slowly convergent. In this paper, we propose a
many-body perturbation correction procedure for the excitation cross section
calculations, taking all singly and doubly excited correlations into account,
but treating their effects on excitation cross sections approximately. We show
that the results agree well with \citet{gchen07} and
\citet{gchen08b} for the 3C and 3D cross sections, and all \ion{Fe}{17} line
ratios are brought into agreement with the \textit{Chandra} observations.

\section{Review of existing theoretical calculations}
Theoretical study of \ion{Fe}{17} X-ray lines and their applications in
astronomy have a long history. \citet{bely67} obtained the excitation cross
sections of $2p^5 3l$ configurations using the Coulomb-Born
approximation. \citet{loulergue73, loulergue75} developed collisional
radiative models of \ion{Fe}{17} and \ion{Ni}{19} with DW collision strengths,
and compared the resulting line intensities with solar
observations. \citet{smith85}, \citet{raymond86}, and \citet{goldstein89}
pointed out the 
importance of resonant excitation and its role in determining the
temperature dependence of \ion{Fe}{17} line ratios. \citet{bhatia92} presented
a complete set of \ion{Fe}{17} line intensities obtained with the DW
approximation. \citet{cornille94} studied the \ion{Fe}{16} satellite transitions
that blend with \ion{Fe}{17} lines. \citet{mohan97} investigated the effects
of relativistic term coupling effects of the \ion{Fe}{17} collision strengths
using the Breit-Pauli $R$-matrix approximation.

\cite{beiersdorfer94} built a flat crystal spectrometer on the electron beam
ion trap (EBIT) of the Lawrence Livermore National Laboratory (LLNL), and
conducted the first laboratory measurement of \ion{Fe}{17} X-ray line
emission. \citet{brown98} carried out a systematic investigation of the
\ion{Fe}{17} lines with the same spectrometer and EBIT facility, and showed
that most previous theories significantly 
underestimate the 3D/3C ratio. \citet{beiersdorfer02,beiersdorfer04}
showed that the $3s$/$3d$ ratio is also underestimated in theory by large 
factors as compared with both EBIT and Tokamak measurements. Independent
laboratory measurements by \citet{laming00} with the National Institute of
Technology and Standards (NIST) EBIT obtained 3D/3C ratios that are consistent with
those from the LLNL EBIT, but reported $3s$/$3d$ ratios very different from
those measured at the LLNL EBIT and concluded that the $3s$/$3d$ ratios are
consistent with modern theoretical calculations. These laboratory work
prompted new theoretical studies to resolve these
discrepancies. \citet{bhatia03} improved the earlier work of \citet{bhatia92}
by including more configurations and computing collision strengths in a wider
energy range. \citet{doron02} investigated the effects of resonant
excitation, recombination of \ion{Fe}{18} and inner-shell ionization of
\ion{Fe}{16} as line formation processes using the Hebrew University Lawrence
Livermore Atomic Code (HULLAC), and concluded that such a multi-ion
approach increases the $3s$/$3d$ ratios, bringing them in better agreements
with laboratory measurements and astrophysical observations. \citet{gu03}
reached similar conclusions and also showed that recombination processes are
important for other iron ions. \citet{gchen02}, \citet{gchen03} and
\citet{gchen05} carried out
a large scale Breit-Pauli $R$-matrix calculation for electron collisional
excitation of \ion{Fe}{17}, and claimed that their results resolve the
discrepancies between theories, laboratory measurements, and astrophysical
observations. An independent close-coupling calculation with
the Dirac atomic $R$-matrix code \citep{loch06} confirms some of the findings
in \citet{gchen03}, but found 
remaining discrepancies between theories and measurements. Most recently,
\citet{gchen07} and \citet{gchen08a} revised their earlier work by including
pseudo states and the $3d^2$ pair excitation configuration in the target
wavefunciton expansion, and obtained significantly lower 3C cross sections than
\citet{gchen03}. 

\begin{figure}
\plotone{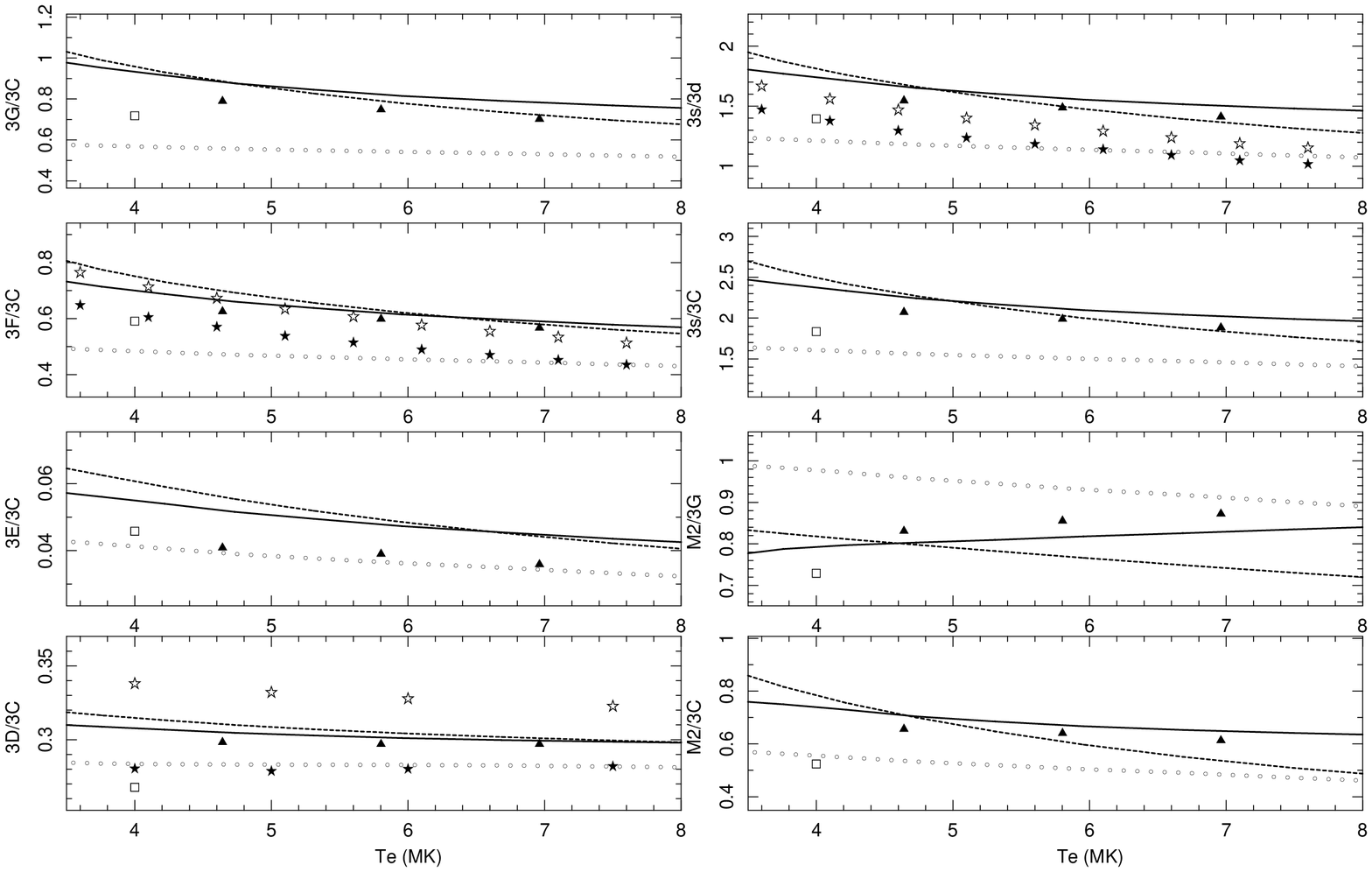}
\caption{\label{fig:tr}Comparison of \ion{Fe}{17} line ratios in various
  theories. The solid line are those of \citet{gu03}; broken lines are those
  of \citet{loch06}, open circles are from APEC, filled triangles are those of
  \citet{doron02}, open squares are from \citet{bhatia03}, filled stars
  are those of \citet{gchen03} for 3D/3C and 3F/3C ratio, and \citet{gchen05}
  for $3s$/$3d$ ratio, the open stars are those of \citet{gchen07} for the
  3D/3C ratio. For 3F/3C and $3s$/$3d$, the open stars are those of
  \citet{gchen03} and \citet{gchen05} scaled for the updated 3C and 3D cross
  sections of \citet{gchen07}.}
\end{figure}

In the present paper, we concentrate on these newer theoretical work. A
comparison of the \ion{Fe}{17} line ratios from these calculations are shown
in Figure \ref{fig:tr}, along with the predictions from the widely used
astrophysical 
plasma emission code (APEC) of \citet{smith01}, which is based on DW collisional
data without resonances. All line ratios involved are calculated at the low
electron density limit of $\le 10^{10}$~cm$^{-3}$.
It is immediately clear from this comparison that the
predictions from APEC and \citet{bhatia03} are very different from those of
\citet{doron02}, \citet{gu03}, and \citet{loch06}. These differences can be
attributed to the lack of resonant excitation contributions in APEC and
\citet{bhatia03}. On the other hand, the results of \citet{doron02} and
\citet{gu03} are relatively close to each other both in the temperature
dependence of the line ratios and their magnitudes, which is not surprising
since the two calculations 
employ essentially the same physics and approximations. The relatively large
difference in the 3E/3C ratio between \citet{doron02} and \citet{gu03} can be
attributed to the incomplete treatment of resonant excitation in
\citet{doron02}. The results of \citet{gu03} and \citet{loch06} are also in
very good agreement, although \citet{loch06} generally produce a steeper
temperature dependence of the line ratios, especially for the $3s/3d$ ratios. This
is due to the fact that \citet{loch06} do not include recombination
contributions to the line intensities. Recombination preferentially enhances
the $3s$ transitions and has larger contributions at higher temperatures due
to the higher fractional abundance of \ion{Fe}{18} relative to
\ion{Fe}{17}. One surprising fact in this comparison is the relatively large
differences between the 3F/3C ratios of \citet{gchen03}, 3s/3d ratios of
\citet{gchen05}, and the corresponding ratios of \citet{loch06}. In fact, the
ratios of \citet{gchen03} and \citet{gchen05}, which include resonant
excitation, are similar to the DW values without resonant
excitation. \citet{loch06} showed that the differences in the 3C and 3D cross
sections in the two $R$-matrix calculations are small. Therefore the large
differences we note in this comparison are due to the differences in the $3s$
intensities. The updated 3C and 3D cross sections of \citet{gchen07} give
larger 3D/3C ratios than all other calculations. Because the full results of
the new calculations of \citet{gchen07} are not available, we scaled the 3F/3C
and $3s$/$3d$ ratios of \citet{gchen03} and \citet{gchen05} to reflect the
updated 3C and 3D cross sections, but keeping $3s$ intensities
unchanged. These scaled ratios for 3F/3C become close to those of \citet{gu03}
and \citet{loch06}, but those for $3s$/$3d$ are still significantly smaller.

\section{\textit{Chandra} stellar observations of \ion{Fe}{17} spectra}
\label{sec:hetg}
In order to assess the accuracies of various calculations, we compare them
with an extensive set of stellar coronal observations with the HETG spectrometer
aboard \textit{Chandra} X-ray observatory. Since its launch, \textit{Chandra}
has observed a large sample of stellar coronal sources with the high resolution
HETG spectrometer. Table \ref{tab:src} lists the sources in our selection. The data
were retrieved from the \textit{Chandra} archive, and reprocessed with CIAO
version 3.4 and CALDB version 3.4. Some sources are observed multiple times,
and we combine the individual observations to improve the statistics. There
are 11 sources that do not provide decent counting statistics
individually, and we have also combined them to produce a single composite
observation referred to as ``Stack'' in Table \ref{tab:src}. Of the 24
targets (after grouping of multiple observations and weak sources) in our
sample, Capella provides the best statistical quality because it is an
calibration target, and observed many times.

\begin{figure}
\plotone{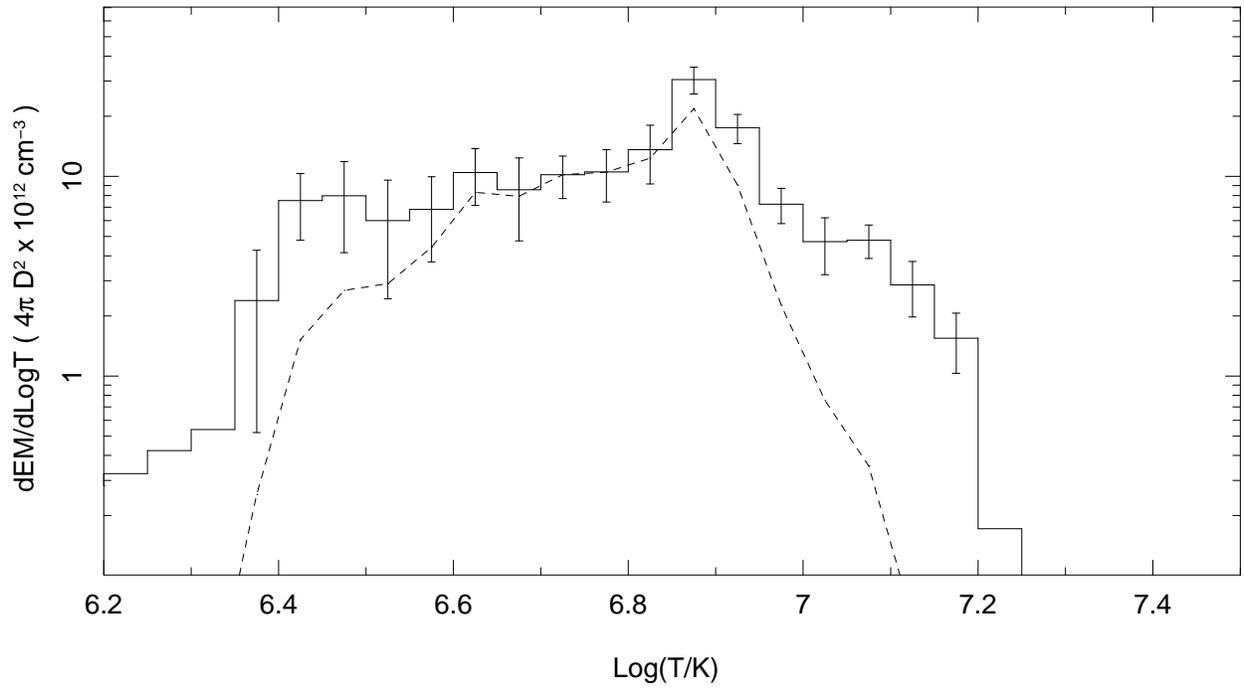}
\caption{\label{fig:dem}Reconstructed DEM of the
  combined Capella observation. $D$ in the unit is the distance of the source
  in cm. The broken line show the total \ion{Fe}{17} emissivity weighted DEM.}
\end{figure}

\begin{figure}
\plotone{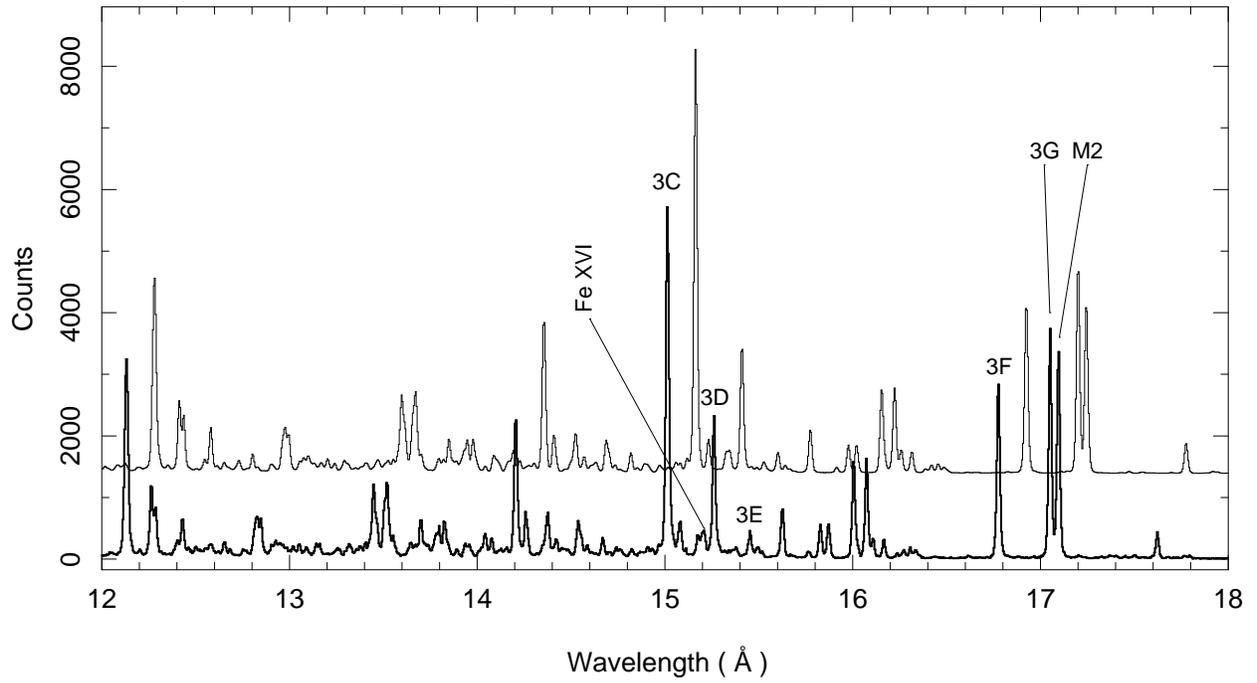}
\caption{\label{fig:sp}\textit{Chandra} HETG spectrum of the combined
  observation of Capella in the 12--18~{\AA}. The lower curve is the sum of
  $\pm$1 orders of MEG 
  data, the upper curve is the model fit with the reconstructed DEM and
  elemental abundances, which is shifted to the upper-right for clarity. The
  \ion{Fe}{17} lines and the \ion{Fe}{16} line at 15.21~{\AA} are marked.}
\end{figure}

We follow the differential emission measure (DEM) analysis method of
\citet{gu06} to characterize the coronal property of these sources. Using the
APEC database modified by the line emissivities of \citet{gu03} and wavelengths
of \citet{gu05}, we reconstruct the DEM distribution and elemental abundances
of the coronal plasma by jointly fitting the $\pm$1 orders of medium and high
energy grating (MEG and HEG) spectra. The resulting DEM and comparison of
theoretical and observed spectra in the 12--18~{\AA} region is shown in
Figures \ref{fig:dem} and \ref{fig:sp} for the composite observation of
Capella. The DEM is peaked near $10^{6.9}$~K, which is similar to the result of
\citet{gu06} using a single observation. To characterize the temperature
of the \ion{Fe}{17} emission regions, it is more appropriate to examine the
DEM weighted by the \ion{Fe}{17} emissivities, which is also shown in
Figure~\ref{fig:dem}. It indicates that the \ion{Fe}{17} emissivity weighted
DEM has a narrower profile. In fact, the characteristic
temperatures of the \ion{Fe}{17} emission regions are always in the range of
$10^{6.6}$--$10^{6.9}$~K for all sources, even though some of them have
significant high temperature tails beyond $10^7$~K. This is due to the
fact that \ion{Fe}{17} fractional abundance peaks near $10^{6.6}$~K in
collisional ionization equilibrium.

The spectral model shown in Figure~\ref{fig:sp} already provides a reasonable
fit to the observed spectrum. However, close inspection of the \ion{Fe}{17}
lines reveal significant discrepancies between them. Using the reconstructed
spectrum as a basis, we obtain a better fit to the \ion{Fe}{17} emission by
varying 3C, 3D, 3E, 3G, 3F, and M2 line intensities. \citet{brown01} and
\citet{brickhouse06} demonstrated that the 3D line blends perfectly with an
\ion{Fe}{16} satellite transition, and another nearby \ion{Fe}{16} transition
at 15.21~{\AA} can be used to remove the \ion{Fe}{16} contribution to the 3D
line. Many of the sources in our sample indicate the presence of the
15.21~{\AA} line, providing a measure of contamination to the 3D line. The
15.21~{\AA} line intensity is extracted from the measured spectra and
multiplied by a factor of 0.7, which is then subtracted from the apparent 3D
intensity. The branching ratio of 0.7 used here is determined with the Flexible
Atomic Code developed by the author, and is between the values 0.83 used by
\citet{brickhouse06} and 0.51 used by \citet{brown01}. However, because the
\ion{Fe}{16} contamination of the 3D line is relatively small, the correction
decreases the measured 3D/3C ratio only by $\sim$~10\%. The differences in the
branching ratios leads to 2--3~\% differences in the 3D/3C ratio, which are
smaller than the statistical errors for most sources. The derived line ratios
for the 24 sources are listed in Table~\ref{tab:ratio}. The \ion{Fe}{17}
emissivity weighted average temperature, $T_{\mbox{em}}$, of the coronae are
also shown. \citet{ness03} reported the ratios involving the 3C, 3D and 3F
lines for a large sample of coronal sources observed by \textit{Chandra} and
\textit{XMM-Newton}. The results obtained here are generally consistent with
those of \citet{ness03} for the set of common sources.

\begin{figure}
\plotone{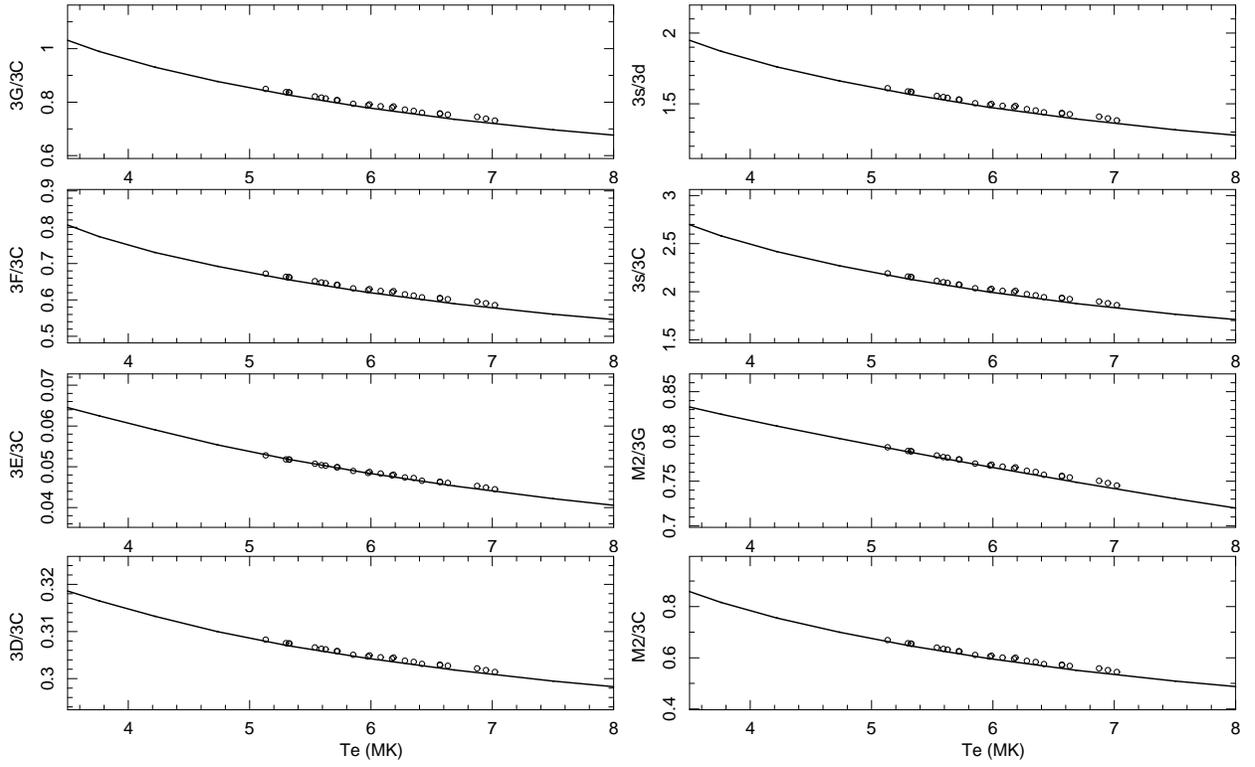}
\caption{\label{fig:wr}Comparison of \ion{Fe}{17} line ratios calculated with
  the reconstructed DEM of the observations (solid lines), and those
  calculated at a single temperature of $T_{\mbox{em}}$ (open circles).}
\end{figure}

In comparing the measured and theoretical line ratios, the calculations should
ideally be performed for the appropriate DEM distributions. In practice, we
find that the ratios calculated with the DEM is very close to the ones
calculated at a single temperature of $T_{\mbox{em}}$. This is demonstrated in
Figure~\ref{fig:wr} for the 8 line ratios we investigate. In the rest of the
paper, we therefore use the isothermal line ratios calculated at
$T_{\mbox{em}}$ for comparison. 

\begin{figure}
\plotone{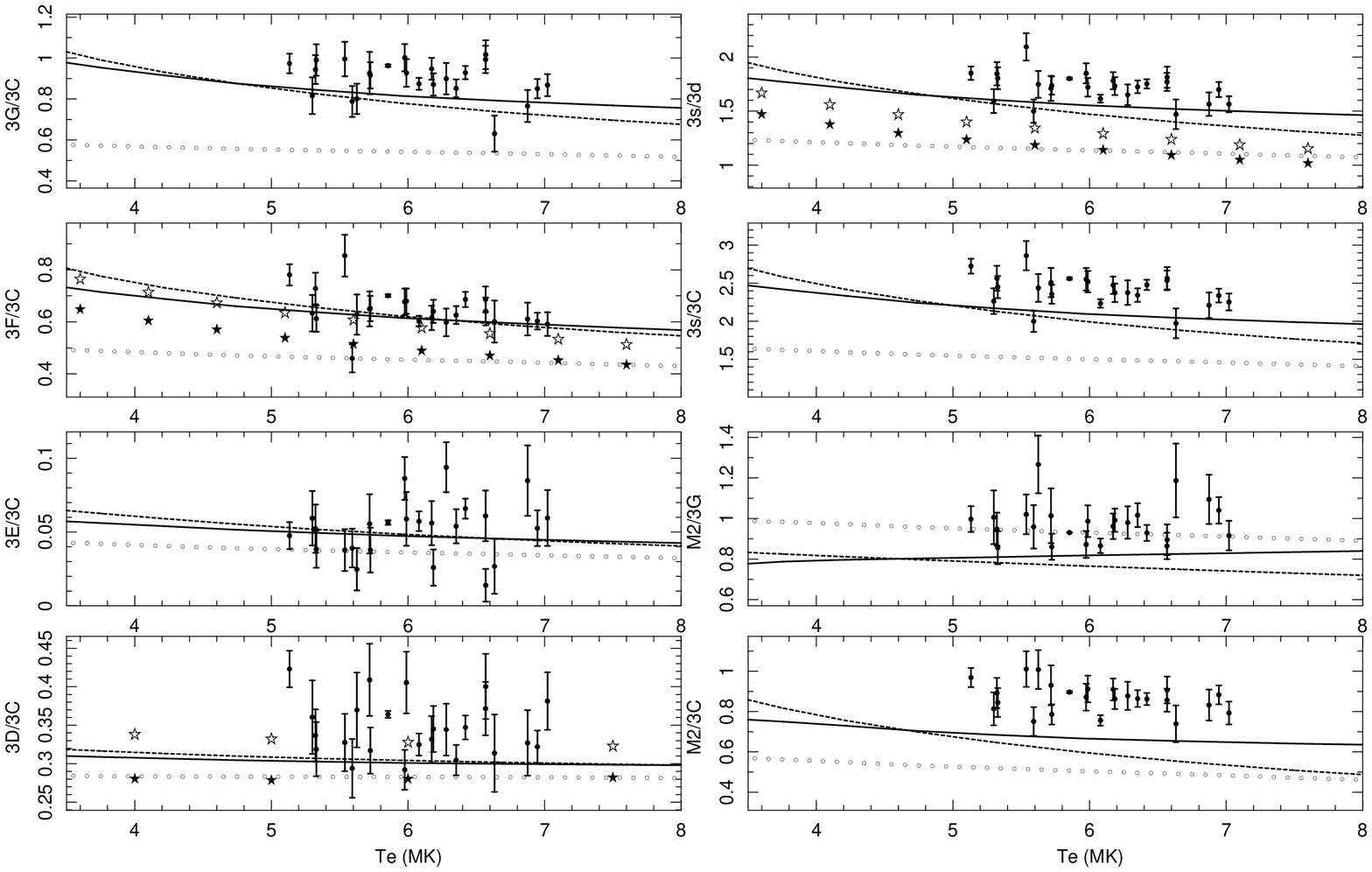}
\caption{\label{fig:mr}Comparison of measured and predicted \ion{Fe}{17} line
  ratios. Filled circles with error bars are the measured values, solid lines
  are those of \citet{gu03}, broken lines are those of \citet{loch06}, open
  circles are from APEC, , filled stars
  are those of \citet{gchen03} for 3D/3C and 3F/3C ratio, and \citet{gchen05}
  for $3s$/$3d$ ratio, the open stars are those of \citet{gchen07} for the
  3D/3C ratio. For 3F/3C and $3s$/$3d$, the open stars are those of
  \citet{gchen03} and \citet{gchen05} scaled for the updated 3C and 3D cross
  sections of \citet{gchen07}.}
\end{figure}

Figure~\ref{fig:mr} shows the comparison of measured and theoretical line
ratios in detail. Because the predictions of \citet{doron02} are generally
similar to those of \citet{gu03}, and those of \citet{bhatia03} are similar to
those of APEC, we omit those from Figure~\ref{fig:mr} to reduce
clutter. It is clear that none of the theories can explain all line
ratios satisfactorily. All theoretical ratios are below the measured ones,
except for 3D/3C, where the results of \citet{gchen07} agree with the
measurements, 3E/3C and 3F/3C, where the values of \citet{loch06}, 
\citet{gu03}, and the results of \citet{gchen03} scaled for the new 3C cross
sections of \citet{gchen07} agree with the measurements, and M2/3G, where the
values of APEC agree with the measurements. However, it should be pointed out
that the measured 
3E/3C ratio have large statistical uncertainties. The APEC
ratios of M2/3C and 3G/3C are very different from the measurements, and the
agreement for M2/3G is therefore fortuitous. Despite the claim of excellent
agreements between their theoretical line ratios, laboratory measurements and
astrophysical observations in \citet{gchen02}, \citet{gchen03}, and
\citet{gchen05}, we find their theoretical values are significantly smaller
than the \textit{Chandra} measurements for the 3F/3C and $3s$/$3d$ ratios,
even more so than the calculations of \citet{loch06} and \citet{gu03}. 
After scaling for the new lower 3C cross sections of \citet{gchen07}, the
$3s$/$3d$ ratios are still smaller than the \textit{Chandra} observations. Since
the data from \citet{gchen03} and subsequent work are not publicly available,
we are not able to calculate other line ratios for thermal plasmas for a more
detailed comparison. 

\section{Distorted-wave with many-body perturbation correction}
The fact that the calculations of \citet{gu03} and \citet{loch06} agree with
each other very well except for the differences in the recombination
contributions indicates that the two theoretical methods, DW in \citet{gu03}
and $R$-matrix in \citet{loch06} are of comparable quality for
\ion{Fe}{17}. The differences between measured and calculated line ratios are
not due to deficiencies in the scattering approximations. The cross section
measurements for 3C and 3D lines of \citet{brown06} demonstrated that
theories tend to overestimate the 3C cross sections more than 3D. The general
trend in the comparison of theories and \textit{Chandra} measurements indeed
support this picture. If the theoretical 3C cross sections were to decrease
by $\sim$~20\%, the predictions of \citet{gu03} and \citet{loch06} would agree
with the observations much better. 3C is a dipole allowed transition, its
collisional excitation cross section is nearly proportional to the oscillator
strength, which is determined solely by atomic structure
calculations. Therefore we suspect that the root problem in the previous
theories of \ion{Fe}{17} line intensities is the incomplete treatment of electron
correlation effects in the target ion. This conclusion is supported by the
comparison of 
various oscillator strengths calculations for 3C and 3D listed in Table~2 of
\citet{gchen03}. Of these calculations, the many-body perturbation theory 
(MBPT) treatment of \citet{safronova01} is likely to be the most accurate in
describing electron correlation effects in \ion{Fe}{17}, which also yields
the smallest oscillator strength for the 3C transition. The recent
calculations of \citet{gchen07} and \citet{gchen08a} also showed that by
improving the target wavefunction expansion, the 3C cross sections are reduced
by 15-20\%. However, it is not clear that the target description in
\citet{gchen07} and \citet{gchen08a} has fully converged, given that the
largest change comes with the inclusion of the $3d^2$ pair excited
configuration, and such pair correlations are know to be notoriously slowly
convergent.

We have developed an independent MBPT method within the framework of FAC for
general open-shell ions, and have used it in calculating the wavelengths of
iron and nickel L-shell X-ray transitions \citep{gu05}. We recently extended
the method to also calculate radiative transition rates and oscillator
strengths. The 3C and 3D weighted oscillator strengths calculated in our MBPT
method are 2.17 and 0.62, respectively. This compares to 2.50 and 0.62
calculated by the DW approximation. These independent MBPT calculations
therefore suggest that the 3C cross sections are likely to be overestimated by
as much as $\sim$~10--20 \% in previous non-MBPT calculations. \citet{gchen07}
gave the weighted oscillator strengths of 3C and 3D to be 2.25 and 0.635,
respectively, which agree with our MBPT results very well. This indicates the
near convergence of their accounting of \ion{Fe}{17} target correlation effects.

However, the extension of the MBPT method to treat collisional excitation processes
is rather difficult beyond the plane-wave Born approximation. Here we propose
an approximate procedure for including MBPT corrections in the calculation of
DW excitation cross sections. \citet{barshalom88} showed that the DW collision
strengths can be factorized to contain angular and radial integrals as follows
\begin{equation}
\label{eq_cs}
\Omega_{01} = 2\sum_{k}\sum_{\alpha_0\alpha_1\atop\beta_0\beta_1}
Q^k(\alpha_0\alpha_1;\beta_0\beta_1)
<\psi_0||Z^k(\alpha_0,\alpha_1)||\psi_1>
<\psi_0||Z^k(\beta_0,\beta_1)||\psi_1>,
\end{equation}
where $\alpha$, $\beta$ denote bound orbitals making the transition, $Z^k$ are
angular factors, and $Q^k$ are radial integrals involving partial-wave sum
\begin{equation}
\label{eq_Qk}
Q^k(\alpha_0\alpha_1;\beta_0\beta_1) = \sum_{\kappa_0\kappa_1}[k]^{-1}
P^k(\kappa_0\kappa_1;\alpha_0\alpha_1)P^k(\kappa_0\kappa_1;\beta_0\beta_1),
\end{equation}
and 
\begin{equation}
\label{eq_Pk}
P^k(\kappa_0\kappa_1;\alpha_0\alpha_1)=
X^k(\alpha_0\kappa_0;\alpha_1\kappa_1)+\sum_{t}
(-1)^{k+t}[k]\sixj{j_{\alpha_0}}{j_1}{t}{j_0}{j_{\alpha_1}}{k}
X^t(\alpha_0\kappa_0;\kappa_1\alpha_1),
\end{equation}
where $\kappa$ represents a continuum orbital and $X^k$ are the usual Slater
integrals. The first term in $P^k$ is the direct contribution, and the second
term is the exchange contribution. In the Coulomb-Bethe approximation, the
direct term can be written as
\begin{equation}
\label{eq_PkApprox}
P^k_{CB}(\kappa_0\kappa_1;\alpha_0\alpha_1) = 
M_k(\alpha_0\alpha_1)R_k(\kappa_0\kappa_1),
\end{equation}
where 
\begin{eqnarray}
M_k(\alpha_0\alpha_1) &=& <\alpha_0||C^k||\alpha_1>
\int (P_{\alpha_0}P_{\alpha_1}+Q_{\alpha_0}Q_{\alpha_1})r^kd r \nonumber\\*
R_k(\kappa_0\kappa_1) &=&
\int (P_{\kappa_0}P_{\kappa_1}+Q_{\kappa_0}Q_{\kappa_1})
\frac{1}{r^{k+1}}d r,
\end{eqnarray}
where $C^k$ is the normalized spherical harmonic, $P_{\alpha}$ and
$Q_{\alpha}$ are the large and small components of the Dirac radial
wavefunctions of orbital $\alpha$. Note that $M_k$ depends only on the target
orbitals and is proportional to the electric multipole integral of rank
$k$. This integral is the same as the one used in radiative transition rate
calculations. Therefore, the ratio of the MBPT corrected to the conventional
DW direct excitation contributions is the same as the ratio of the MBPT
corrected to uncorrected oscillator strenghts. We calculate such ratios from the
radiative transition rates with and without MBPT effects, and use them as
scaling factors to correct the direct excitation contributions to 
$P^k$, and obtain the MBPT corrected collision strengths in the DW
approximation. 

\begin{figure}
\plotone{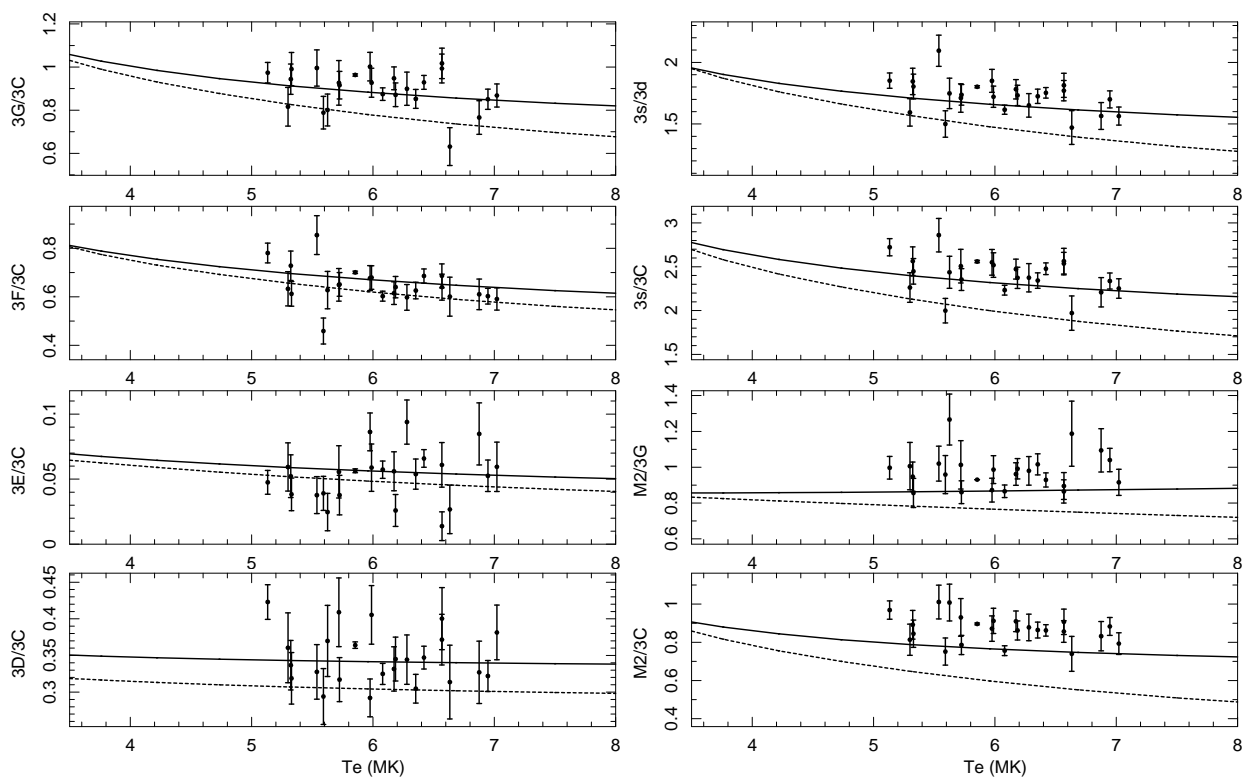}
\caption{\label{fig:nr}Comparison of measured \ion{Fe}{17} line ratios with
  the predictions of the present MBPT corrected theory (solid lines). The
  predictions of \citet{loch06} are also shown as broken curves.}
\end{figure}

Following the procedure outlined above, and the three-ion model of
\citet{gu03}, we recalculate the \ion{Fe}{17} line ratios and compare them
with the \textit{Chandra} observations in Figure~\ref{fig:nr}. The results of
\citet{loch06} are shown again for easy comparison. It is obvious that our
MBPT corrections bring the calculated line ratios into much better agreement
with \textit{Chandra} observations. The M2/3C and M2/3G ratios seem to be still
slightly below the observed ones, but show significant improvements over the
existing theories.

\begin{figure}
\plotone{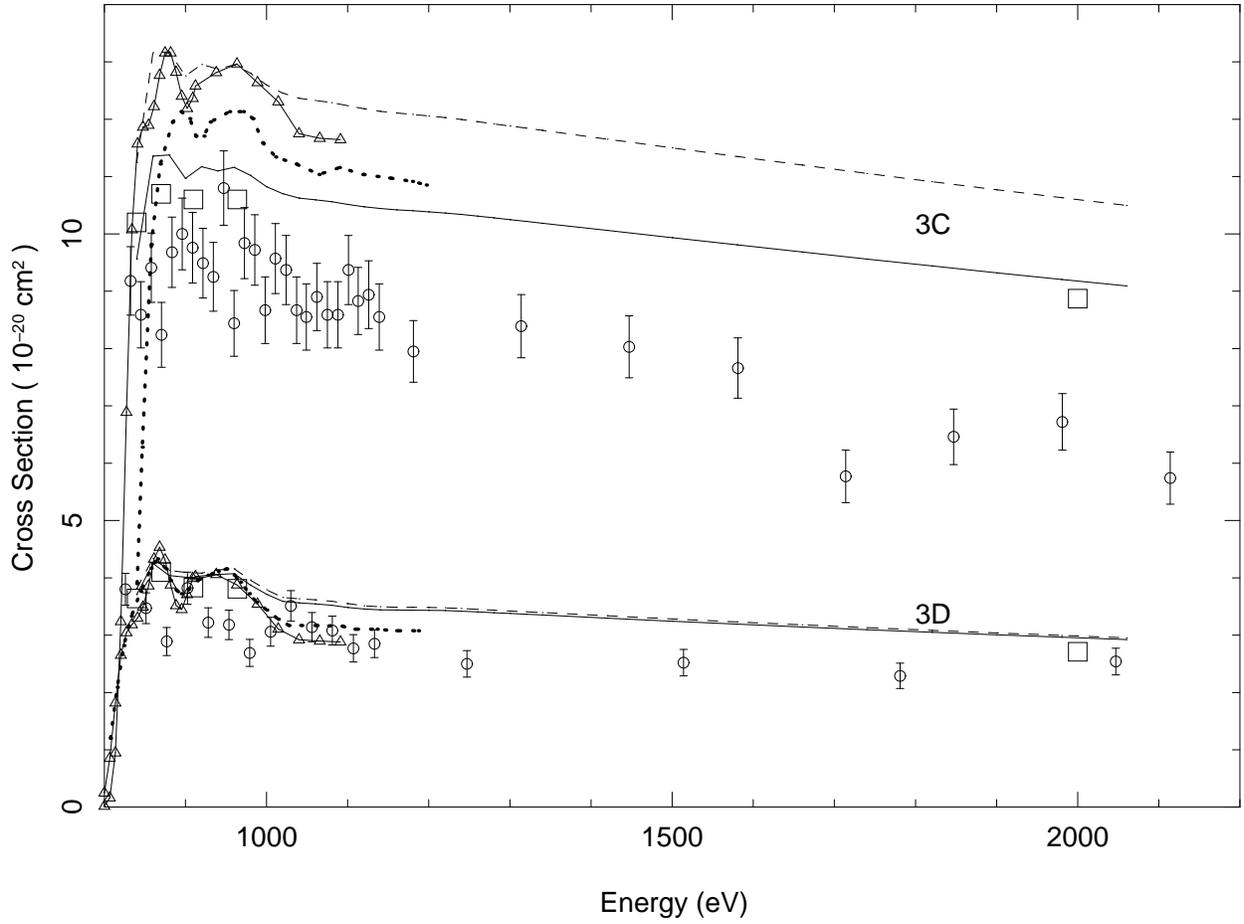}
\caption{\label{fig:cross}Comparison of measured and predicted 3C and 3D cross
  sections. The open circles with error bars are the laboratory measurements
  of \citet{brown06}, solid lines with no symbols are the present MBPT
  corrected calculation, dashed lines are the values from \citet{gu03}, dotted
  lines are those of \citet{loch06}, solid lines with filled triangles are
  those of \citet{gchen03}, and open squares are the calculations of
  \citet{gchen07}.}  
\end{figure}

Finally, we compare the MBPT corrected effective formation cross sections of
the 3C and 3D transitions with laboratory measurements of
\citet{brown06} in Figure~\ref{fig:cross}. Our effective cross sections
include contributions of resonant excitation and radiative cascades from
higher levels. The cross sections of \citet{gu03}, \citet{gchen03},
\citet{loch06}, and \citet{gchen07} are also shown for comparison. The
$R$-matrix calculations 
include resonances, but do not taken into account radiative cascades. Clearly,
the MBPT corrections leave the 3D cross sections of \citet{gu03} almost
unchanged, but decrease the 3C cross sections by about 18\%. The cross
sections of \citet{gchen03} and \citet{loch06} are all higher than the present
MBPT corrected values for 3C. The cross sections of \citet{gchen07} are only
slightly lower than the present calculations by $\sim$5\%, which likely
represents the differences between different scattering
approximations. However, the present MBPT corrected cross sections for both 3C
and 3D are still about 15\% higher than the laboratory measurements overall,
while the 
3D/3C ratios agree with each other very well. The energy dependence of the
measured cross sections shown in Figure~\ref{fig:cross} were obtained by
sweeping the electron beam energies and recording the energy dependence of the
3C and 3D line intensities. They are normalized to the RR emission into the
M-shell of \ion{Fe}{16} at a single energy point of 964~eV. There is
significant statistical fluctuation in the measured data, but the overall
discrepancies between the measured values and the present calculations are
less than 15\%, especially at the near threshold region below 1200~eV. The
discrepancies between the measured values and the results of \citet{gchen07}
are even smaller, at 10\% level. If the suggestion that the RR cross sections
used in \citet{brown06} are underestimated by 25\% \citep{gchen08b} is
correct, the new theoretical cross sections of 3C and 3D of the present paper
and those of \citet{gchen07} and \citet{gchen08a} would have become
smaller than the renormalized measurements.

Since our new \ion{Fe}{17} line ratios agree with \textit{Chandra}
observations, the lower experimental 3C and 3D cross sections would lead to
lower cross sections for all other lines as well, and to inaccurate iron abundance
determinations if these lines are used as main constraints. The cross sections
of \citet{brown06} were determined by normalizing the collisional excitation
lines to the RR emission onto the M-shell of
\ion{Fe}{16}. The overall normalization uncertainties were estimated to be on 
the order of 10\%.  The differences between our new calculations and the
measurements are therefore not significantly larger than the experimental
uncertainties, especially if one takes into account that the more elaboate
R-Matrix scattering method seems to further lower the cross sections by
$\sim$5\%. 

\section{Density diagnostics with the M2 line}
Due to the relatively slow radiative decay of the M2 transition, the M2 line
is collisionally quenched at electron densities above $10^{13}$~cm$^{-3}$. The
M2/3G ratio has 
been used to infer electron densities in magnetic cataclysmic variables
\citep{mauche01} and stellar coronal sources \citep{ness05}. However,
\S\ref{sec:hetg} shows that the M2/3G ratios of all coronal sources with
densities known to be below $10^{12}$~cm$^{-3}$ are significantly larger than
previous theoretical predictions, which is identified as a major problem
in applying this diagnostic in \citet{ness05}. We note that \citet{ness05}
seem to indicate that the low density limit of M2/3G calculated with
\citet{gu03} data agree with the coronal observations. We have since confirmed
that it is due to some misinterpretation of the \citet{gu03} data in
calculating the M2/3G ratio, and the \citet{gu03} ratio of M2/3G is in fact
much lower than the observations. The present MBPT corrected M2/3G ratio, although
appears to be still slightly lower overall, is at least marginally consistent
with low density coronal observations.

\begin{figure}
\plotone{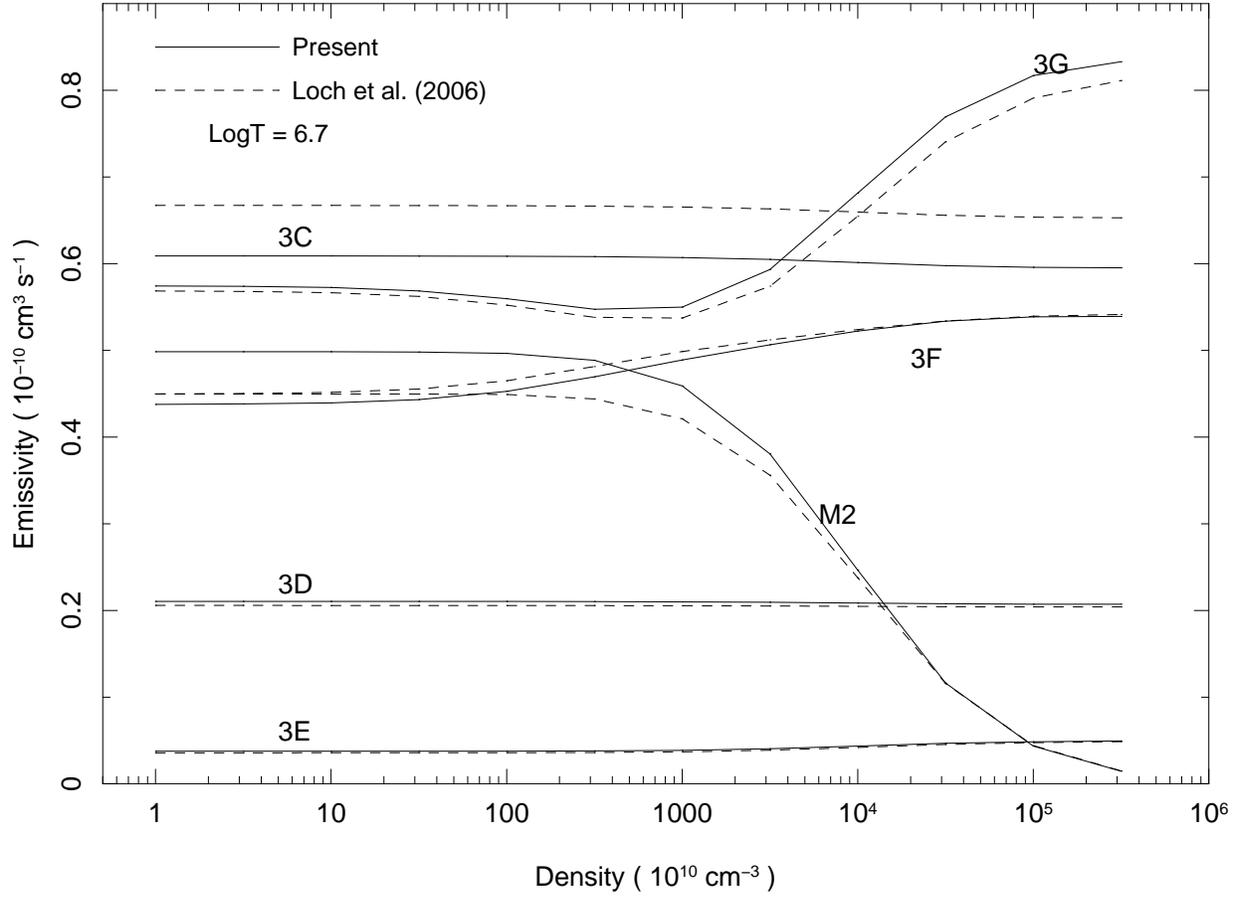}
\caption{\label{fig:emd}Density dependence of the \ion{Fe}{17} line emissivity
  at a temperature of $10^{6.7}$~K. The solid lines are the present MBPT
  corrected calculations, and dashed lines are those of \citet{loch06}.}
\end{figure}

\begin{figure}
\plotone{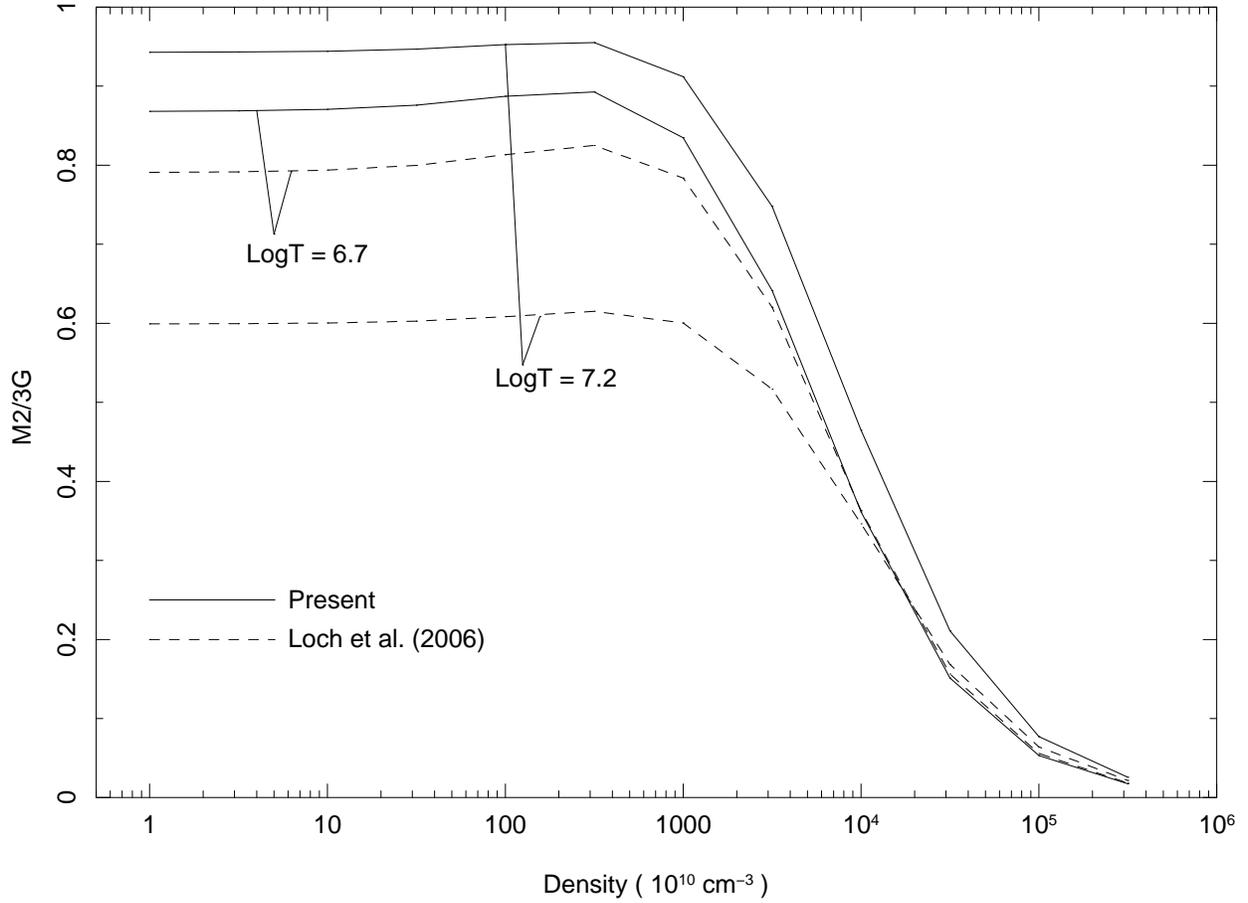}
\caption{\label{fig:emr}Density dependence of the M2/3G line ratio at two
  marked electron temperatures. The solid lines are the present MBPT
  corrected calculations, and dashed lines are those of \citet{loch06}.}
\end{figure}

In Figure~\ref{fig:emd}, we show the density dependence of the emissivities of
the six \ion{Fe}{17} lines at a temperature of $10^{6.7}$~K in comparison with
the predictions of \citet{loch06}. The publicly released data from
\citet{loch06} do not include forbidden transition rates. In order to derive
the density dependence of line emissivities, we augmented their dataset with
the forbidden transition rates from the present calculations. Two major
differences are seen between the present work and that of \citet{loch06},
namely, the smaller 3C (due to MBPT correction) and larger M2 (due to
recombination contributions) in the present calculation, which helps to
explain the observed coronal values. The density dependence of M2/3G ratios in
the present calculation and those of \citet{loch06} are shown in
Figure~\ref{fig:emr} for two electron temperatures, $10^{6.7}$ and
$10^{7.2}$~K. The differences between the present work and \citet{loch06} are
even more pronounced at higher temperatures due to the increasing importance
of recombination contributions to the M2 line intensity.

\section{conclusions}
We have conducted a systematic study of the \ion{Fe}{17} emission lines of
stellar coronal sources observed by the HETG spectrometer aboard
\textit{Chandra}. At least some of the line ratios involving the six principal
\ion{Fe}{17} lines are shown to disagree with most existing theories. The
calculations of \citet{doron02} and \citet{gu03}, which include resonant
excitation and recombination contributions in DW approximation provide as good
agreement with data as the more sophisticated close-coupling calculation of
 \citet{loch06} with the Dirac atomic $R$-matrix code. Similarly sophisticated,
Breit-Pauli $R$-matrix results of \citet{gchen02,gchen03,gchen05}, however, show
worse agreements with the \textit{Chandra} observations for the 3F/3C and
$3s$/$3d$ line ratios. The source of discrepancies in the two $R$-matrix
calculations are not clear. The most recent Breit-Pauli $R$-matrix results of
\citet{gchen07} gave the best agreement with laboratory measurements of 3C and
3D cross sections, but a systematic comparison with 
\textit{Chandra} line ratios cannot be carried out, as the full results of
these calculations necessary for collisional radiative modeling are not
available. If we assume that the $3s$ line intensities of \citet{gchen07} are
the same as \citet{gchen03} and \citet{gchen05}, and only correct the ratios
for the updated 3C and 3D cross sections, the resulting 3D/3C and 3F/3C ratios
are in agreement with the \textit{Chandra} observations, but the $3s$/$3d$
ratios are still significantly lower than observed values.

Based on the MBPT calculations of oscillator strengths,
we suggest that the incomplete treatment of electron correlation effects of
the \ion{Fe}{17} target is the cause of overestimating the 3C cross
sections in most previous theories, as is also suggested by the recent
calculations of \citet{gchen07} and \citet{gchen08a}. We propose a simple,
MBPT based procedure to 
incorporate such correlation effects in the calculation of electron
collisional excitation cross sections under DW approximation. The resulting
\ion{Fe}{17} line ratios are shown to agree with \textit{Chandra} observations
much better than previous predictions. The 3C and 3D cross sections of the
present work are only $\sim$5\% larger than the calculations of
\citet{gchen07}. Comparison of our MBPT corrected 3C
and 3D cross sections and the measurements of \citet{brown06} suggests
remaining discrepancies on the order of 15\%, although systematic
uncertainties of about 10\% may explain part of this difference, and the
simplied scattering approximation used here may be responsible for the
remaining discrepancies. We have shown that if the RR cross sections used for 
normalization in \citet{brown06} are underestimated by 25\%, as suggested by
\citet{gchen08b}, the new theoretical cross sections of the present work and
those of \citet{gchen07} and \citet{gchen08a} would have become smaller than
the renormalized measurements.

We have presented the density dependence of the \ion{Fe}{17} emissivities and M2/3G
ratio, and compared them with previous calculations. The improved low density
limit of M2/3G ratio provides greater confidence in applying this density
diagnostics in astrophysical observations.

\acknowledgements
The authour would like to thank P. Beiersdorfer for extensive comments on the
original manuscript. Support for this work was provided by the National
Aeronautics and Space Administration through Chandra Award Number TM7-8001Z
issued by the Chandra 
X-ray Observatory Center, which is operated by the Smithsonian Astrophysical
Observatory for and on behalf of the National Aeronautics Space Administration
under contract NAS8-03060. This work was performed under the auspices of the
U.S. Department of Energy by Lawrence Livermore National Laboratory under
Contract DE-AC52-07NA27344.


\begin{deluxetable}{cccc} 
\tabletypesize{\scriptsize}
\tablecaption{\label{tab:src}\textit{Chandra} HETG observations selected in
  the present sample. Targets with multiple exposures are combined to form
  a single observation. 11 weak sources are also grouped to form a composite
  observation for target 15, or ``Stack''.}
\tablehead{ 
\colhead{Index} & 
\colhead{Target} & 
\colhead{ObsId} & 
\colhead{Exposure (ks)} 
}
\startdata
     1 &        $\xi$ Uma &   1894 &   70.9 \\
     2 &           AD Leo &   2570 &   45.2 \\
     3 &       $\tau$ Sco &    638 &   59.2 \\
     4 &           EV Lac &   1885 &  100.0 \\
     5 &           CC Eri &   4513 &   89.5 \\
     6 &         HD 45348 &    636 &   94.6 \\
     7 &           AU Mic &     17 &   58.8 \\
     8 &           44 Boo &     14 &   59.1 \\
     9 &           VW Cep &   3766 &  116.6 \\
    10\tablenotemark{a} &          Capella & $\cdots$ &  297.9 \\
$\cdots$ &          Capella &     57 &   28.8 \\
$\cdots$ &          Capella &   1099 &   14.6 \\
$\cdots$ &          Capella &   1100 &   14.6 \\
$\cdots$ &          Capella &   1101 &   14.6 \\
$\cdots$ &          Capella &   1103 &   40.5 \\
$\cdots$ &          Capella &   1235 &   14.6 \\
$\cdots$ &          Capella &   1236 &   14.6 \\
$\cdots$ &          Capella &   1237 &   14.6 \\
$\cdots$ &          Capella &   1318 &   26.7 \\
$\cdots$ &          Capella &   2583 &   27.6 \\
$\cdots$ &          Capella &   3674 &   28.7 \\
$\cdots$ &          Capella &   5955 &   28.7 \\
$\cdots$ &          Capella &   6471 &   29.6 \\
    11 &         HD 93497 &   3410 &   57.0 \\
    12 &           AB Dor &     16 &   52.3 \\
    13 &           TZ Crb &     15 &   83.7 \\
    14 &         V824 Ara &   2538 &   94.2 \\
    15\tablenotemark{b} &            Stack & $\cdots$ &  711.7 \\
$\cdots$ &           TY Pyx &    601 &   49.1 \\
$\cdots$ &           FK Com &    614 &   41.4 \\
$\cdots$ &         HD 68273 &    629 &   64.9 \\
$\cdots$ &           II Peg &   1451 &   42.7 \\
$\cdots$ &        HD 206267 &   1888 &   34.1 \\
$\cdots$ &        HD 206267 &   1889 &   39.5 \\
$\cdots$ &        HD 111812 &   1891 &  130.2 \\
$\cdots$ &        HD 223460 &   1892 &   95.7 \\
$\cdots$ &      PROXIMA Cen &   2388 &   42.4 \\
$\cdots$ &        HIP 92680 &   3729 &   73.9 \\
$\cdots$ &        V4046 Sgr &   5423 &   97.7 \\
    16 &           ER Vul &   1887 &  112.0 \\
    17 &          HR 1099 &  62538 &   94.7 \\
    18 &     $\beta$ Ceti &    974 &   86.1 \\
    19 &            Algol &    604 &   51.7 \\
    20\tablenotemark{c} &           AR Lac & $\cdots$ &   64.3 \\
$\cdots$ &           AR Lac &      6 &   32.1 \\
$\cdots$ &           AR Lac &      9 &   32.2 \\
    21 &           UX Ari &    605 &   48.5 \\
    22\tablenotemark{d} &           IM Peg & $\cdots$ &  192.1 \\
$\cdots$ &           IM Peg &   2527 &   24.6 \\
$\cdots$ &           IM Peg &   2528 &   24.8 \\
$\cdots$ &           IM Peg &   2529 &   24.8 \\
$\cdots$ &           IM Peg &   2530 &   23.9 \\
$\cdots$ &           IM Peg &   2531 &   23.9 \\
$\cdots$ &           IM Peg &   2532 &   22.5 \\
$\cdots$ &           IM Peg &   2533 &   23.7 \\
$\cdots$ &           IM Peg &   2534 &   23.9 \\
    23\tablenotemark{e} &     $\sigma$ Gem & $\cdots$ &  120.8 \\
$\cdots$ &     $\sigma$ Gem &   5422 &   62.8 \\
$\cdots$ &     $\sigma$ Gem &   6282 &   57.9 \\
    24 &    $\lambda$ And &    609 &   81.9 \\
\enddata

\tablenotetext{a}{Sum of the following 13 observations of Capella.}
\tablenotetext{b}{Sum of the following 11 observations of different targets.}
\tablenotetext{c}{Sum of the following 2 observations of AR Lac.}
\tablenotetext{d}{Sum of the following 8 observations of IM Peg.}
\tablenotetext{e}{Sum of the following 2 observations of $\sigma$ Gem.}

\end{deluxetable}
\clearpage

\begin{deluxetable}{ccccccccccc}
\tabletypesize{\scriptsize}
\rotate
\tablecaption{\label{tab:ratio}\ion{Fe}{17} line ratios measured with the
  \textit{Chandra} HETG observations. $T_{\mbox{em}}$ is the \ion{Fe}{17}
  emissivity weighted average temperature of sources in units of $10^6$~K. The
  numbers in parentheses are statistical uncertainties. $3s$ refers to the sum
  of 3F, 3G, and M2, and $3d$ refers to the sum of 3C, 3D, and 3E.}
\tablehead{
\colhead{Index} & 
\colhead{Target} & 
\colhead{$T_{\mbox{em}}$ (MK)} & 
\colhead{3D/3C} & 
\colhead{3E/3C} & 
\colhead{3F/3C} & 
\colhead{3G/3C} & 
\colhead{M2/3C} & 
\colhead{M2/3G} & 
\colhead{3s/3C} & 
\colhead{3s/3d}
}
\startdata
 1 &        $\xi$ Uma & 5.13 & 0.423(0.024) & 0.048(0.009) & 0.780(0.041) & 0.974(0.048) & 0.969(0.048) & 0.997(0.063) & 2.723(0.098) & 1.852(0.062)\\
 2 &           AD Leo & 5.30 & 0.360(0.048) & 0.059(0.018) & 0.633(0.070) & 0.816(0.089) & 0.814(0.082) & 1.006(0.132) & 2.263(0.168) & 1.594(0.111)\\
 3 &       $\tau$ Sco & 5.32 & 0.337(0.034) & 0.052(0.016) & 0.728(0.060) & 0.944(0.070) & 0.892(0.075) & 0.947(0.081) & 2.564(0.163) & 1.846(0.107)\\
 4 &           EV Lac & 5.33 & 0.319(0.035) & 0.038(0.013) & 0.612(0.051) & 0.991(0.077) & 0.845(0.072) & 0.856(0.082) & 2.448(0.145) & 1.804(0.101)\\
 5 &           CC Eri & 5.54 & 0.328(0.037) & 0.038(0.014) & 0.854(0.080) & 0.996(0.084) & 1.011(0.088) & 1.020(0.098) & 2.861(0.192) & 2.095(0.126)\\
 6 &         HD 45348 & 5.59 & 0.294(0.038) & 0.039(0.013) & 0.459(0.054) & 0.788(0.076) & 0.751(0.071) & 0.959(0.107) & 1.999(0.140) & 1.500(0.108)\\
 7 &           AU Mic & 5.63 & 0.370(0.049) & 0.025(0.014) & 0.628(0.077) & 0.801(0.074) & 1.008(0.096) & 1.266(0.142) & 2.437(0.183) & 1.748(0.124)\\
 8 &           44 Boo & 5.72 & 0.317(0.030) & 0.038(0.015) & 0.651(0.049) & 0.916(0.064) & 0.786(0.050) & 0.861(0.064) & 2.354(0.124) & 1.737(0.084)\\
 9 &           VW Cep & 5.72 & 0.409(0.047) & 0.056(0.020) & 0.649(0.067) & 0.927(0.103) & 0.930(0.099) & 1.013(0.135) & 2.507(0.192) & 1.711(0.115)\\
10 &          Capella & 5.85 & 0.364(0.004) & 0.056(0.002) & 0.701(0.005) & 0.963(0.005) & 0.896(0.005) & 0.930(0.000) & 2.560(0.015) & 1.802(0.009)\\
11 &         HD 93497 & 5.98 & 0.292(0.026) & 0.086(0.015) & 0.677(0.051) & 1.002(0.066) & 0.872(0.067) & 0.872(0.066) & 2.551(0.147) & 1.850(0.093)\\
12 &           AB Dor & 5.99 & 0.405(0.040) & 0.059(0.018) & 0.679(0.049) & 0.927(0.067) & 0.912(0.066) & 0.987(0.078) & 2.519(0.140) & 1.721(0.086)\\
13 &           TZ Crb & 6.08 & 0.325(0.014) & 0.057(0.007) & 0.603(0.021) & 0.875(0.029) & 0.756(0.026) & 0.866(0.035) & 2.233(0.056) & 1.616(0.036)\\
14 &         V824 Ara & 6.17 & 0.332(0.030) & 0.056(0.015) & 0.615(0.047) & 0.948(0.053) & 0.910(0.054) & 0.962(0.063) & 2.473(0.114) & 1.782(0.079)\\
15 &            Stack & 6.19 & 0.345(0.030) & 0.026(0.012) & 0.640(0.044) & 0.871(0.055) & 0.862(0.050) & 0.991(0.057) & 2.374(0.121) & 1.732(0.083)\\
16 &           ER Vul & 6.28 & 0.344(0.034) & 0.094(0.017) & 0.598(0.053) & 0.899(0.077) & 0.878(0.070) & 0.980(0.080) & 2.375(0.161) & 1.651(0.096)\\
17 &          HR 1099 & 6.35 & 0.305(0.020) & 0.054(0.011) & 0.626(0.034) & 0.853(0.044) & 0.864(0.042) & 1.016(0.059) & 2.343(0.088) & 1.725(0.059)\\
18 &     $\beta$ Ceti & 6.42 & 0.347(0.016) & 0.066(0.007) & 0.686(0.029) & 0.929(0.032) & 0.862(0.030) & 0.929(0.040) & 2.477(0.067) & 1.753(0.043)\\
19 &            Algol & 6.57 & 0.372(0.035) & 0.061(0.017) & 0.686(0.049) & 0.993(0.067) & 0.857(0.057) & 0.865(0.065) & 2.536(0.129) & 1.771(0.083)\\
20 &           AR Lac & 6.57 & 0.400(0.042) & 0.014(0.011) & 0.640(0.053) & 1.017(0.070) & 0.907(0.066) & 0.895(0.075) & 2.564(0.147) & 1.814(0.098)\\
21 &           UX Ari & 6.63 & 0.314(0.050) & 0.027(0.019) & 0.601(0.080) & 0.631(0.088) & 0.740(0.091) & 1.187(0.182) & 1.971(0.196) & 1.471(0.137)\\
22 &           IM Peg & 6.88 & 0.327(0.043) & 0.085(0.024) & 0.610(0.063) & 0.766(0.078) & 0.832(0.077) & 1.094(0.121) & 2.209(0.168) & 1.565(0.109)\\
23 &     $\sigma$ Gem & 6.95 & 0.322(0.021) & 0.053(0.012) & 0.603(0.032) & 0.851(0.047) & 0.883(0.047) & 1.040(0.065) & 2.337(0.092) & 1.700(0.069)\\
24 &    $\lambda$ And & 7.02 & 0.381(0.037) & 0.059(0.019) & 0.591(0.046) & 0.868(0.053) & 0.793(0.057) & 0.916(0.073) & 2.252(0.111) & 1.564(0.073)\\
\enddata
\end{deluxetable}
\clearpage

\clearpage

\clearpage

\clearpage

\clearpage

\clearpage

\clearpage

\clearpage

\clearpage

\end{document}